# Mott switching and structural transition in the metal phase of VO$_2$ nanodomain[*]

Chang-Yong Kim[1,*], Tetiana Slusar[2,*], Jinchul Cho[2,3], Hyun-Tak Kim[2,3]

[1]Canadian Light Source 44 Innovation Boulevard Saskatoon, SK S7N 2V3, Canada

[2]Metal-Insulator-Transition Laboratory, Electronics and Telecommunications Research Institute, Daejeon 34129, Republic of Korea

[3]Department of Advanced Device Technology, University of Science and Technology, Daejeon 34113, Republic of Korea

[*]These authors contributed equally: Chang-Yong Kim, Tetiana Slusar

**ABSTRACT:** VO$_2$ undergoes the insulator-metal transition (IMT) and monoclinic-rutile structural phase transition (SPT) near 67°C. The IMT switching has many applications. However, there is an unresolved issue whether the IMT is a Mott transition or a Peierls transition. This complication is caused by metal and insulator coexistence, which is an inherent property of the IMT region. Thus, the acquired data in the IMT region are averaged over the two phases in many experiments. We overcome the issue by probing the electronic state of the monoclinic structure and by introducing a model that accounts for the coexisting phases. We reveal the Mott IMT in the non-distorted monoclinic nanodomain between 55-63°C, and the distortion-assisted SPT above 60°C.

**INTRODUCTION:** Thermally induced first-order transition in vanadium dioxide (VO$_2$) at $T_c \approx$ 67℃ involves changes of both electronic and lattice structures from a low temperature monoclinic (M marking M1) insulator to a high temperature tetragonal rutile (R) metal.[1] The exact cause of the transition has been unknown and intensely debated for the past several decades. In particular, two competing mechanisms have been introduced as essential to induce the insulator-metal transition (IMT or electronic switching) in VO$_2$: the Peierls structural instability that implies formation/disappearance of the tilted vanadium dimers[2]-[3] and the Mott switching that involves a sudden increase/decrease in the magnitude of the critical Coulomb interaction between electrons[4]-[5]. Moreover, the combined Peierls-Mott scenarios have also been suggested.[6]-[7]

On the other hand, there are numerous works suggesting that the electronic and structural transitions do not occur simultaneously.[8]-[18] Their separation has been achieved by observations of the monoclinic metal phase (MMP), which unambiguously proves the Mott transition occurring prior to the M-to-R structural change. Nevertheless, the uncertainty in understanding the driving mechanism of the IMT still exist. It is attributed to the formation of the mixed phase (insulator and metal) region inherent in the first-order phase transition. This complicates the characterization of the individual phases, thus, hindering the description of the intrinsic IMT pathway.[8],[19]

One way to reveal the IMT mechanism in VO$_2$ is to trace only one phase throughout the transition. For example, nanoscale infrared spectroscopy at ≈ 10 μm wavelength has been known for its capability to sense the metallic phase very effectively throughout the whole transition regime.[20] In this paper, we explore the electronic state of the selected monoclinic structure through measuring the X-ray absorption spectra by diffraction anomalous near-edge structure (DANES)[21]-[24] experiments that use a synchrotron light source. DANES probes the electronic structure of a material by sensing the modulation of the diffraction intensities across the absorption edge of a selected V-atom. The X-ray diffraction intensity is proportional to the absolute square of the structure factor which is determined by the spatial arrangements of constituent atoms weighted by the atomic form factors. When the incident X-ray energy changes across absorption edge of the constitute atom, the atomic form factor is modulated significantly. X-ray absorption spectroscopy can be seen as energy dependent X-ray diffraction with momentum transfer $Q = 0$ (straight through a sample). Thus, DANES is able to reveal the electronic state of the specific structure, which is an essential to clarify the complex phase-transition pathway.

This approach allows us to experimentally manifest the metal phase that emerges in the non-distorted monoclinic structure (i.e., Mott transition) on heating and to observe the further monoclinic distortion by means of the monoclinic nanodomain's size reduction due to the rutile phase nucleation. Moreover, we theoretically support the observed Mott-type IMT in VO$_2$ by deducing the "Mottian" characteristics of the MMP, such as the divergence of the Brinkman-Rice effective mass and large correlation strength. Also, based on the extended Brinkman-Rice picture,[8],[19] we succeeded in developing the mixed phase model able to distinguish and quantify the properties of the coexisting phases in the transition region. Thus, it enables us to observe the evolution of the individual phases in the vicinity of the IMT in VO$_2$.

Figure 1a shows the resistance vs temperature curve measured upon heating of a VO$_2$ film on the r-plane Al$_2$O$_3$ substrate. The curve demonstrates a noticeable deviation from the Arrhenius law (solid blue line) and an electronic transition characteristic between $T_{c(onset)}^{IMT} \approx 55°C$ and $T_{c(end)}^{IMT} \approx 63°C$ with resistance drop of almost four orders of magnitude. Figures 1b and 1c depict the evolution of the off-axis diffraction peaks obtained by measuring the in-situ synchrotron-based X-ray diffraction (XRD) on heating. Here, the left peak gradually disappears while the center and the right peaks quite maintain their intensities. The left peak can be assigned to the monoclinic $(11\bar{1})_M$ peak, the center peak to $(011)_M$ from one of the twin domains, and the right side peak to $(011)_M$ from the other twin domain (see details in the Supporting Information).[25] Although the $(011)_M$ and $(11\bar{1})_M$ peaks from the monoclinic structure disappear with an increasing temperature, the rutile phase $(101)_R$ peaks simultaneously develop close to the $(011)_M$ peaks. Because of a broad peak width, $(011)_M$ and $(101)_R$ are not resolved as separated peaks. Hence, a gradual increase of $(101)_R$ concomitant with a decrease of $(011)_M$ is apparent from the shift of peak positions across the structural phase transition (SPT), as clearly shown between 65-70°C in Figure 1c.

To examine the SPT, we selected the $(11\bar{1})_M$ peak (Figures 1b and 1c), which is unique to the monoclinic structure, and traced its changes on heating. Figure 1d shows the progression of the full width at the half maximum (FWHM) of $(11\bar{1})_M$ (see inset of Figure 1d) throughout the phase transition, which is plotted on the same x-axis scale as the resistance curve in Figure 1a for a convenient comparison of the two processes. Notably, the temperature difference between $T_{c(onset)}^{IMT} \approx 55°C$ and the starting temperature of the monoclinic structure distortion (a shrinkage of a monoclinic nanodomain) $T_{c(onset)}^{FWHM} > 60°C$ (red solid line in Figure 1d) is about 5°C (Figures 1a and 1d). This observation implies the existence of the monoclinic metal phase (MMP) between $55 - 60°C$ and, thus, the isostructural Mott IMT. Above 60°C (Figure 1d), the peak broadening is observed due to the distortion of the monoclinic structure. We assume that this distortion is caused by the local emergence (nucleation) of the rutile metal phase (RMP) that gradually spreads upon further heating until the SPT is complete above 70°C (Figure 1c).

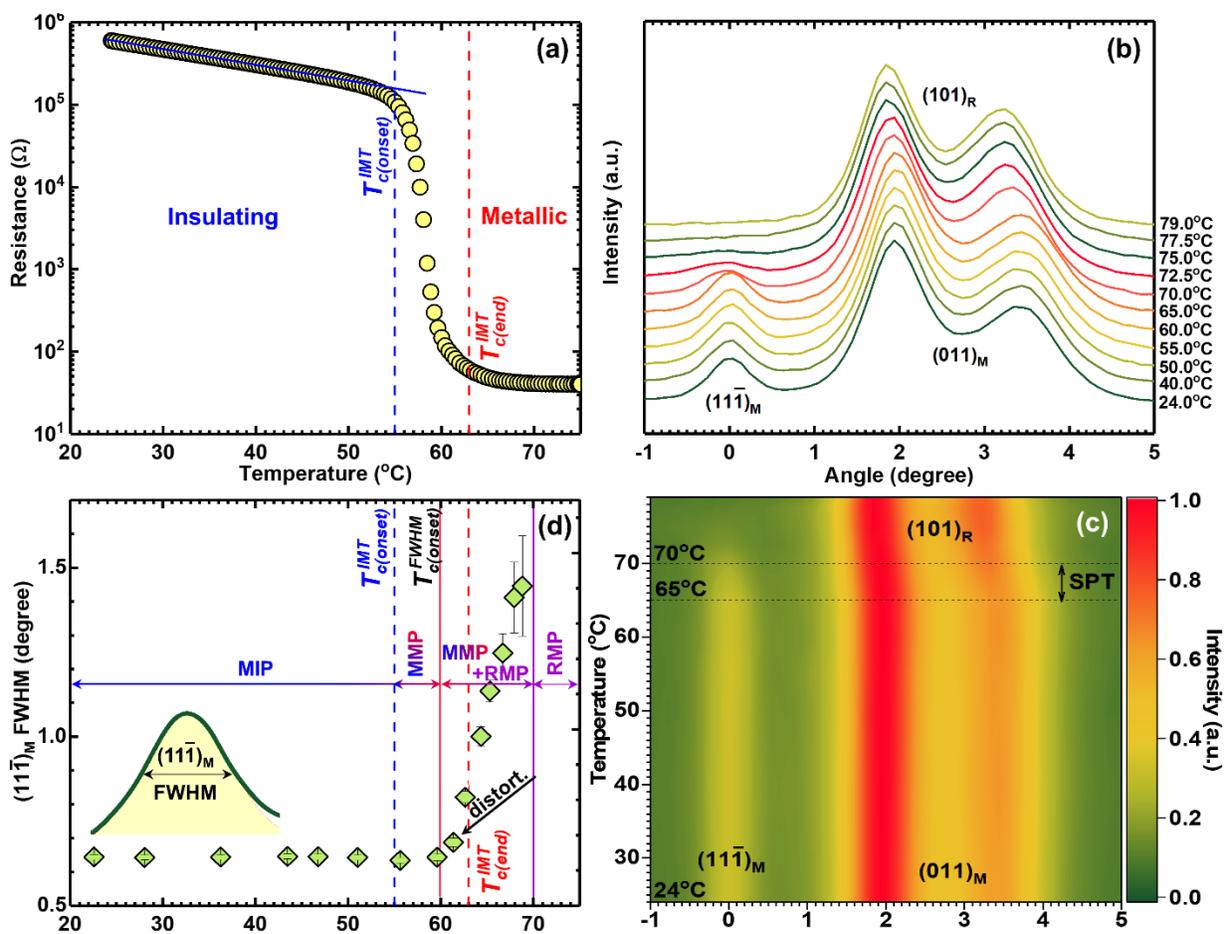

**Figure 1.** Evolution of the electronic and crystalline structures of VO$_2$ on heating. (a) The insulator-to-metal transition (IMT) revealed by a sudden resistance ($R$) drop between $T_{c(onset)}^{IMT} \approx$ 55°C and $T_{c(end)}^{IMT} \approx 63°C$. (b, c) Evolution of the off-axis XRD peaks from the monoclinic (M) and rutile (R) VO$_2$. (d) Changes of the FWHM of the $(11\bar{1})_M$ peak (inset) on heating denoting the following regions: the monoclinic metal phase (MMP) region between $55 - 60°C$ emerged from the monoclinic insulator phase (MIP) as a result of the Mott transition; nucleation of the rutile metal phase (RMP) at $T_{c(onset)}^{FWHM} > 60°C$ revealed by the monoclinic distortion (increase of the FWHM) and, thus, the MMP+RMP region between $60 - 70°C$; the RMP region formed by the SPT above 70°C.

Figure 2a shows the two representative DANES profiles obtained by measuring the intensities of the monoclinic $(200)/(\bar{2}11)_M$ peak at room temperature and the rutile $(101)_R$ peak at 75°C, while changing the incident X-ray energy across the V K edge. As inferred by comparing the profiles, a significant difference can be found in the pre-edge region around 5468 eV, as marked by the black arrows in Figure 2a. In particular, a dip in diffraction intensity is flattened in the metallic rutile phase compared to the insulating monoclinic. A similar observation has been previously reported for XANES measurements.[26] Therefore, the pre-edge region is clear evidence of the electronic structure change.

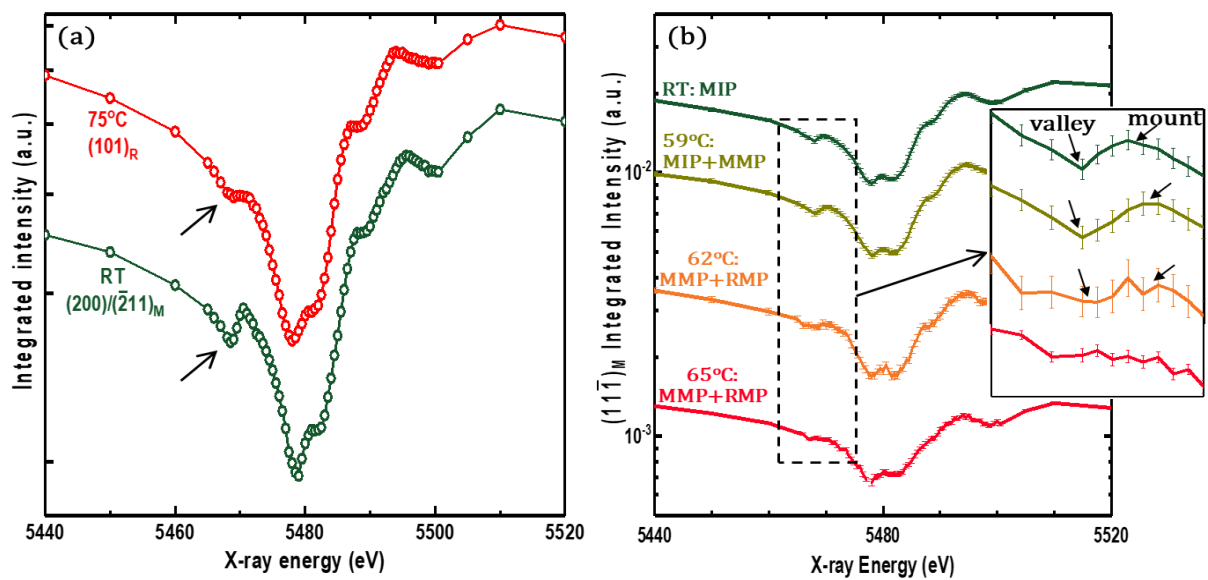

**Figure 2.** DANES at the V K edge from the $VO_2$ film on $Al_2O_3$. (a) Integrated intensities of the diffraction peaks from the monoclinic insulating (RT, $(200)/(\bar{2}11)_M$) and rutile metallic (75°C, $(101)_R$) phases of $VO_2$ vs the incident X-ray energy. Errors in the intensities are smaller than the symbols size. (b) Integrated intensities of the $(11\bar{1})_M$ peak of $VO_2$ as a function of the incident X-ray energy and temperature. Inset: Magnified pre-edge regions from (b) for an X-ray energy range of 5465-5472 eV with designated "valley" and "mount" areas of the distinguishable profile changes observed at 59°C, due to the emergence of the monoclinic metallic phase (MMP) from the monoclinic insulator phase (MIP), and further, above 62°C, due to the nucleation of the rutile metal phase (RMP).

Figure 2b shows the DANES profiles of $(11\bar{1})_M$ taken at various temperatures through the transition. Along with the gradual reduction in the overall signal intensity, we first recognize a subtle, but notable difference in the shape of the "mount" between the DANES profiles taken at room temperature (RT) and 59°C. This indicates the local emergence of the metallic puddles within the non-distorted monoclinic insulating phase (MIP), as observed in Figures 1a and 1d between $55-60°C$. Previously, the metallic puddles have been visualized in the transition regime by the nanoscale infrared spectroscopy.[20] Further, at 62°C and 65°C, both the "valley" and "mount" of the profiles undergo significant changes showing a lower contrast variation similar to that from the metallic rutile phase at 75°C (Figure 2a). This resemblance between the DANES profiles in the monoclinic metallic and rutile metallic phases provides clear evidence that the monoclinic insulating $VO_2$ film has switched to the monoclinic metallic state; this is pivotal for defining the Mott transition in $VO_2$.

We further analyze the detected phases of $VO_2$ on the basis of the obtained experimental information. First of all, we develop an inhomogeneous mixed phase model in order to reveal the metal fraction in the film and apply it to fit the resistance curve in Fig. 1a. The corresponding experimental resistance ratio ($r(T)/(r_0=1250\Omega)$: $r_0$ is a Drude resistance), as shown in Fig. 3a (the solid orange balls), is closely fitted by a calculated $(1/\rho^3)(1-\rho^4)$ curve (open black triangles), as explained in the Supporting Information; the IMT resistance is fitted for the first time. Here $\rho$ (blue curve) is defined as the band-filling factor (band FF)[8],[19] of $0<\rho \leq 1$ and it is extracted from the resistance curve. $\rho$ indicates the fraction of band filling with electrons in $k$-space, and it also indicates the extent of the metal phase in an inhomogeneous system with insulator and metal phases in real-space (see the Supporting Information). The $\rho$ values in k-space and real-space are equivalent. For instance, $\rho = 1$ means half-filling in $k$-space and the electronic structure of metal with one electron per atom in real-space. This reveals that the metal phase exists all over the measurement region. In the case of $\rho=0.5$, the metal phase is 50% in the measurement region (the other 50% is the insulator phase) and the conduction band is filled by one half of half filling as per average over the coexisting phases. A change from $\rho \approx 0.23$ at about 55°C, i.e., the metal phase share in the measurement region is 23% (see the blue diamond in Fig. 3a) and insulator – 77%, to $\rho \approx 0.97$ at 60°C, i.e., nearly metal (the red diamond in Fig. 3a) reveals that the IMT has already occurred at the sharp resistance reduction region.

Further, we estimate the coherent domain size by using the Scherrer formula[27] and the FWHM of the $(11\bar{1})_M$ peak shown in Figure 1d. The details of the calculations are given in the Supporting Information, while the evolution of the monoclinic nanodomains size in the wide temperature range of $23\sim69°C$ is plotted in Figure 3b.

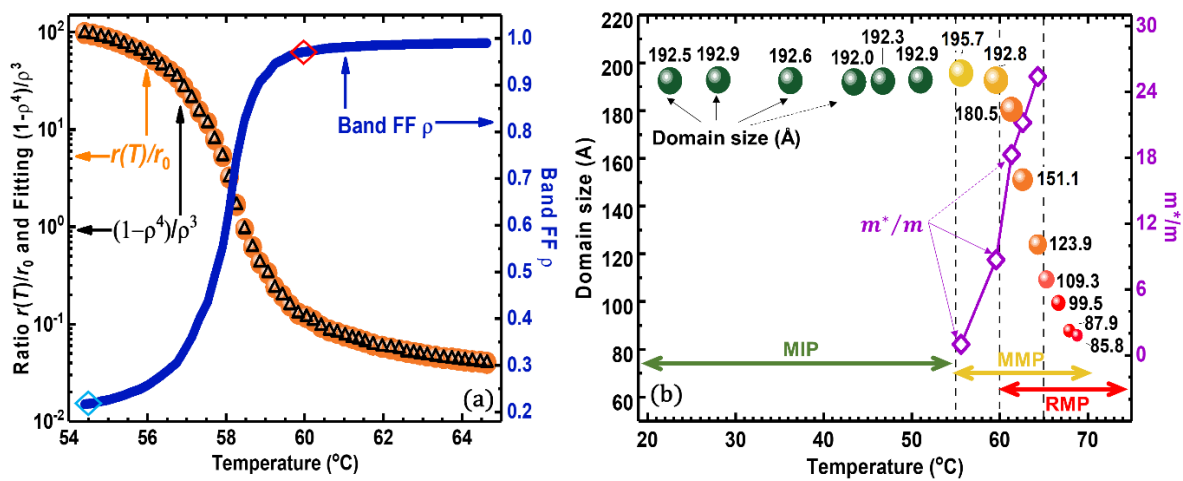

**Figure 3.** Analyzing the resistance and tracking the changes of the $VO_2$ monoclinic nanodomains size upon heating. (a) The temperature dependence of the band-filling factor (band FF or $\rho$, $0 < \rho \leq 1$; blue solid line)[8],[19] extracted from the resistance curve in Fig.1a. The experimental resistance $r(T)/r_0$ curve (orange balls), overlaps with the calculated $(1/\rho^3)(1-\rho^4)$ curve (black open triangles) (details in the Supporting Information). (b) Temperature dependence of the monoclinic domain size (left $y$-axis) and the carriers' effective mass (right $y$-axis). The regions of the monoclinic insulator phase (MIP), monoclinic metal phase (MMP) and rutile metal phase (RMP) and their coexistence are also shown.

In particular, Figure 3b shows the temperature dependence of the sizes of the monoclinic nanodomains, which are approximately the same (within the measurement uncertainty), about 192-195 Å, before and immediately after the IMT. In other words, the domain size conservation is observed in the temperature range of $23\sim60°C$, irrespective of the electronic state switching from the insulator to metal above 55°C. However, when the temperature exceeds $60°C$, the nanodomains start to decrease in size due to the nucleation and growth of the rutile phase.

To estimate the contribution of each individual phase in the coexistence region, we assume the domain-size conservation and set it to be equal to 195.7 Å that corresponds to the domain at 56°C (near the IMT onset). The contribution of the other phases to the 195.7 Å monoclinic domain reduction (domain shrinkage) on heating is taken into account by applying the inhomogeneous mixed phase model with the band filling factor $\rho$ given in Figure 3a. The pie charts in Figure 4 show the resulting share of the MIP, MMP, and RMP phases via the calculated sizes of their nanodomains and relative percentage. The domain sizes along with the $\rho$ values are also given in Table 1.

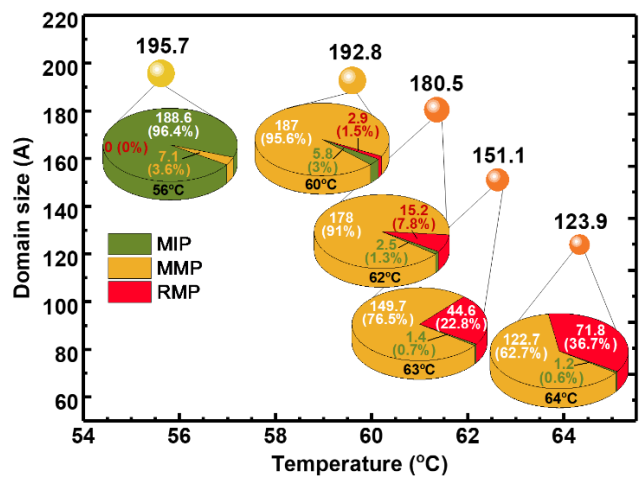

**Figure 4.** The calculated share and percentage of the monoclinic insulating phase (MIP), monoclinic metal phase (MMP) and rutile metal phase (RMP) coexisting within the VO$_2$ nanodomains (extracted from Fig. 3b) in the IMT temperature range 56 ~ 64°C.

It is worth emphasizing the importance of the obtained result regarding the distinction between the coexisting phases and the observation of the Mott transition within the non-distorted monoclinic structure. Note that we associate the observed MIP-MMP transition with the thermal excitations of impurity donor atoms coming from, for example, oxygen deficiency (the indirect IMT). This results in a collapse of the Hubbard $d$-gap in the MIP phase formed by the critical on-site Coulomb repulsive energy $U_c$ between $3d^1$ electrons.[11],[28] The collapse of the $d$-gap induces free carriers by way of the avalanche in the non-distorted MIP. These carriers mutually scatter each other when their concentration tends to the metal one (~ $10^{22}$ cm$^{-3}$). Moreover, we speculate that these scatterings, regarded as correlation, further promote the divergent rise of the effective mass of carriers, $m^*/m$,[8],[19] as shown in Figure 3b.

Further, from the effective mass, we obtain another important IMT characteristic that is the correlation strength, $\kappa_{BR}$, in the Brinkman-Rice (BR) picture.[29] The extended Brinkman-Rice picture applied to inhomogeneous systems[8],[19] reveals the extent of the band-filling factor $\rho$. The corresponding calculations, given in the Supporting Information, determine the correlation strength of $\kappa_{BR} \approx 0.98$ at $\rho \approx 0.99$ at the MMP (near 65°C) and $m^*/m \approx 1/(1-\rho^4) \approx 25$. The large effective mass is consistent with that in a previous paper,[30] which justifies that the extraction of $\rho$ from resistance is correct. Note that although a large correlation strength is obtained it may be incorrect, because the BR picture is applied at zero temperature in the absence of thermal phonons.[8],[19] The analyzed effective masses, the correlation strength, and all of the parameters for their calculations are shown in Table 1.

**Table 1. Temperature dependent values of the band filling factor ($\rho$), domain sizes of different phases (MIP, MMP and RMP), effective mass ($m^*/m$) and correlation strength ($\kappa_{BR}$). Distortion of the structure means the shrinkage (size reduction) of the monoclinic nanodomain.**

| Temp. (°C) | [1]$\rho = n_{carrier}/n_{tot}$[2] | Distortion of the structure | [3]Monoclinic domain size (Å) | [4]MIP size (Å) | [5]MMP size (Å) | [6]RMP (Å) | [7]$m^*/m$ | $\kappa_{BR} = U/U_c$ |
|---|---|---|---|---|---|---|---|---|
| 56 | 0.036 | No | 195.7 | 188.6 | 7.1 | 0 | 1 | [8]not determined at $\rho<1$ |
| 60 | 0.970 | No | 192.8 | 5.8 | 187.0 | 2.9 | 8.7 | 0.941 |
| 62 | 0.986 | Yes | 180.5 | 2.5 | 178.0 | 15.2 | 18.3 | 0.972 |
| 63 | 0.988 | Yes | 151.1 | 1.4 | 149.7 | 44.6 | 21.2 | 0.976 |
| 64 | 0.990 | Yes | 123.9 | 1.2 | 122.7 | 71.8 | 25.4 | 0.980 |
| 66 | n/a | Yes | 109.3 | 0 | 109.3 | 86.4 | n/a | n/a |
| 67 | n/a | Yes | 99.5 | 0 | 99.5 | 96.2 | n/a | n/a |
| 68 | n/a | Yes | 87.9 | 0 | 87.9 | 107.9 | n/a | n/a |
| 69 | n/a | Yes | 85.8 | 0 | 86.8 | 109.9 | n/a | n/a |

**CONCLUSIONS:** By exclusively probing the electronic structure of the monoclinic nanodomains, we demonstrated the Mott transition that occurs via the emergence of the monoclinic metal phase (MMP). This is independent of the structural distortion responsible for the Peierls transition. We also discovered that the size of the monoclinic nanodomain reaches maximum just before rapid reduction, which corresponds to the structural distortion. In addition, the MMP and the RMP are mixed in the distorted region above 60°C, which is due to the structural transition from the MMP to the RMP required to stabilize the metallic phase. Therefore, the extent of the RMP increases with increasing temperature. The MMP is the key in developing numerous applications such as the ultrafast IMT switches, oscillating devices, Mott transistors, optoelectronic devices, neuromorphic devices, IMT qubits etc. Furthermore, we consider that the previously suggested conclusions[(6)-(7)] on the metal-insulator transition in $VO_2$ that is a combination of the electron-electron interaction (Mott transition) and electron-phonon interaction (Peierls transition), are caused by the combined observation of both the distorted monoclinic metal phase and the rutile metal phase in the measurement region.

---

[1] As the temperature increases, the increase of $\rho$ translates into percolation.
[2] $n_{tot}=1.69 \times 10^{22}/cm^3$ is given by the inverse volume of the tetragonal rutile.[8]
[3] Monoclinic domain sizes are obtained from analyzing the X-ray diffraction.
[4] MIP is the monoclinic insulator phase
[5] MMP is the monoclinic metal phase
[6] RMP is the rutile metal phase. MIP, MMP and RMP sizes were obtained by subtracting the domain size from the monoclinic domain size of 195.7Å.
[7] effective mass ($m^*/m$) is defined as following: $m^*/m = 1/(1-\rho^4)$, where $\rho$ is band filling factor, i.e. the extent of the metal phase.
[8] Since the Brinkman-Rice picture is applied at half filling of one electron per atom when $\rho=1$, in the $\rho<1$ case, $\kappa_{BR} = U/U_c$ is not determined.

Our approach to probe the electronic properties of a specific crystallographic phase can be applied to a variety of phase change materials ($NbO_2$, $V_2O_3$, $Fe_3O_4$, InSe etc.) and multiphase systems, thus, paving the way for understanding their intrinsic properties.

**Acknowledgements**
The authors thank to Prof. M. M. Qazilbash and Prof. V. Dobrosavljević for their very valuable comments. The research was financially supported by the National Korean MS&ICT project (2017-0-00830) on MIT. Part of the research described in this paper was performed at the Canadian Light Source, a national research facility of the University of Saskatchewan, which is supported by the Canadian Foundation for Innovation (CFI), the Natural Sciences and Engineering Research Council (NSERC), the National Research Council (NRC), the Canadian Institutes of Health Research (CIHR), the Government of Saskatchewan, and the University of Saskatchewan.

**References**

(1) Morin, F. J. Oxides Which Show a Metal-to-Insulator Transition at the Neel Temperature. *Physical Review Letters* **1959**, 3 (1), 34-36.
(2) Goodenough, J. B. The two components of the crystallographic transition in $VO_2$. *Journal of Solid State Chemistry* **1971**, 3 (4), 490-500.
(3) Wentzcovitch, R. M.; Schulz, W. W.; Allen, P. B. $VO_2$: Peierls or Mott-Hubbard? A view from band theory. *Physical Review Letters* **1994**, 72, 3389-3392.
(4) Mott, N. F. The transition to the metallic state. *The Philosophical Magazine: A Journal of Theoretical Experimental and Applied Physics* **1961**, 6, 287-309.
(5) Rice, T. M.; Launois, H.; Pouget, J. P. Comment on "$VO_2$: Peierls or Mott-Hubbard? A View from Band Theory". *Physical Review Letters* **1994**, 73, 3042.
(6) Paez, G. J.; Singh, C. N.; Wahila, M. J.; Tirpak, K. E.; Quackenbush, N. F.; Sallis, S.; Paik, H.; Liang, Y.; Schlom, D. G.; Lee, T.-L.; Schlueter, C.; Lee, W.-C.; Piper, L. F. J. Simultaneous Structural and Electronic Transitions in Epitaxial $VO_2$/$TiO_2$(001). *Phys. Rev. Lett.* **2020**, 124, 196402.
(7) Lee, D.; Yang, D.; Kim, H.; Kim, J.; Song, S.; Choi, K. S.; Bae, J.-S.; Lee, J.; Lee, J.; Lee, Y.; Yan, J.; Kim, J.; Park, S. Deposition-Temperature-Mediated Selective Phase Transition Mechanism of $VO_2$ Films. *J. Phys. Chem. C* **2020**, 124, 31, 17282–17289.
(8) Kim, H.-T.; Chae, B.-G.; Youn, D.-H.; Maeng, S.-L.; Kim, G.; Kang, K.-Y.; Lim, Y.-S. Mechanism and observation of Mott transition in $VO_2$-based two- and three-terminal devices. *New Journal of Physics* **2004**, 6, 52.
(9) Kim, H.-T.; Lee, Y. W.; Kim, B.-J.; Chae, B.-G.; Yun, S. J.; Kang, K.-Y.; Han, K.-J.; Yee, K.-J.; Lim, Y.-S. Monoclinic and Correlated Metal Phase in $VO_2$ as Evidence of the Mott Transition: Coherent Phonon Analysis. *Physical Review Letters* **2006**, 97, 266401.
(10) Kumar, S.; Strachan, J. P.; Pickett, M. D.; Bratkovsky, A.; Nishi, Y.; William, R. S. Sequential Electronic and Structural Transitions in $VO_2$ Observed Using X-ray Absorption Spectromicroscopy. *Advanced Materials* **2014**, 26, 7505-7509.
(11) Kim, H.-T.; Kim, M.; Sohn, A.; Slusar T.; Seo, G.; Cheong, H.; Kim, D.-W. Photoheat-

induced Schottky nanojunction and indirect Mott transition in VO$_2$: photocurrent analysis. *J. Phys.: Condens. Matter* **2016**, 28, 085602.
(12) Gray, A. X.; Jeong, J.; Aetukuri, N. P.; Granitzka, P.; Chen, Z.; Kukreja, R.; Higley, D.; Chase, T.; Reid, A. H.; Ohldag, H.; Marcus, M. A.; Scholl, A.; Young, A. T.; Doran, A.; Jenkins, C. A.; Shafer, P.; Arenholz, E.; Samant, M. G.; Parkin, S. S. P.; Dürr, H. A. Correlation-Driven Insulator-Metal Transition in Near-Ideal Vanadium Dioxide Films. *Physical Review Letters* **2016**, 116, 116403.
(13) Slusar, T.V.; Cho, J.; Lee, H.; Kim, J.-W.; Yoo, S. J.; Bigot, J.-Y.; Yee, K.-J.; Kim, H.-T. Mott transition in chain structure of strained VO$_2$ films revealed by coherent phonons. *Sci Rep* **2017**, 7, 16038.
(14) Lee, D.; Chung, B.; Shi, Y.; Kim, G.-Y.; Campbell, N.; Xue, F.; Song, K.; Choi, S.-Y.; Podkaminer, J. P.; Kim, T. H.; Ryan, P. J.; Kim, J.-W.; Paudel, T. R.; Kang, J.-H.; Spinuzzi, J. W.; Tenne, D. A.; Tsymbal, E. Y.; Rzchowski, M. S.; Chen, L. Q.; Lee, J.; Eom, C. B. Isostructural metal-insulator transition in VO$_2$. *Science* **2018**, 362, 6418, 1037-1040.
(15) Moatti, A.; Sachan, R.; Cooper, V.; Narayan, J. Electrical Transition in Isostructural VO$_2$ Thin-Film Heterostructures. *Sci Rep* **2019**, 9, 3009.
(16) Shiga, D.; Minohara, M.; Kitamura, M.; Yukawa, R.; Horiba, K.; Kumigashira, H. Emergence of metallic monoclinic states of VO$_2$ films induced by K deposition. *Physical Review B* **2019**, 99, 125120.
(17) Morrison, V. R.; Chatelain, R. P.; Tiwari, K. L.; Hendaoui, A.; Bruhács, A.; Chaker, M.; Siwick, B. J. A photoinduced metal-like phase of monoclinic VO$_2$ revealed by ultrafast electron diffraction. *Science* **2014**, 346, 445.
(18) Grandi, F.; Amaricci, F.; Fabrizio, M., Unraveling the Mott-Peierls intrigue in vanadium dioxide, *Physical Review Research* **2020**, 2, 013298.
(19) Kim, H. T. Analysis of the diverging effective mass in YaBa$_2$Cu$_3$O$_{6+x}$ for high-$T_c$ mechanism and pairing symmetry, *International Journal of Modern Physics* B **2018**, 32, 1840031.
(20) Qazilbash, M. M.; Brehm, M.; Chae, B.-G.; Ho, P.-C.; Andreev, G. O.; Kim, B.-J.; Yun, S. J.; Balatsky, A. V.; Maple, M. B.; Keilmann, F.; Kim, H.-T.; Basov D. N. Mott Transition in VO$_2$ Revealed by Infrared Spectroscopy and Nano-Imaging. *Science* **2007**, 318, 1750-1753.
(21) Stragier, H.; Cross, J. O.; Rehr, J. J.; Sorensen, L. B.; Bouldin, C. E.; Woicik, J. C. Diffraction anomalous fine structure: A new x-ray structural technique. *Physical Review Letters* **1992**, 69, 3064.
(22) Pickering, I. J.; Sansone, M.; Marsh, J.; George, G. N. Diffraction anomalous fine structure: a new technique for probing local atomic environment. *Journal of the American Chemical Society* **1993**, 115, 6302.
(23) Vacinova, J.; Hodeau, J. L.; Wolfers, P.; Lauriat, J. P.; ElKaim, E. Use of Anomalous Diffraction, DAFS and DANES Techniques for Site-Selective Spectroscopy of Complex Oxides. *Journal of Synchrotron Radiation* **1995**, 2, 236.
(24) Kawaguchi, T.; Fukuda, K.; Tokuda, K.; Shimada, K., Ichitsubo, T.; Oishi, M.; Mizuki, J.; Matsubara, E. Revisit to diffraction anomalous fine structure. *Journal of Synchrotron Radiation* **2014**, 21, 1247.
(25) Yang, T.-H.; Aggarwal, R.; Gupta, A.; Zhou, H.; Narayan, R. J. Semiconductor-metal transition characteristics of VO$_2$ thin films grown on c- and r-sapphire substrates. *Journal of Applied Physics* **2010**, 107, 053514.
(26) Marini, C.; Pascarelli, S.; Mathon, O.; Joseph, B.; Malavasi, L.; Postorino, P. Tracking

competitive lattice distortions in strongly correlated VO$_2$-based systems: A temperature-dependent EXAFS study. *EPL (Europhysics Letters)* **2013**, 102, 66004.

(27) Patterson, A. L. The Scherrer Formula for X-Ray Particle Size Determination. *Physical Review* **1939**, 56, 978.

(28) Kim, H. T. *Japan Physical Society Autumn Meeting Abstract* 13aJB-1 **2016**.

(29) Brinkman, W. F.; Rice, T. M. Application of Gutzwiller's variational Method to the Metal-Insulator Transition, *Physical Review* B **1970**, 2, 4302.

(30) Fu, D.; Liu, K.; Tao, T.; Lo, K.; Cheng, C.; Liu, B.; Zhang, R.; Bechtel, H. A.; Wu, J. Comprehensive study of the metal-insulator transition in pulsed laser deposited epitaxial VO$_2$ thin films, *Journal of Applied Physics* **2013**, 113, 043707.

# SUPPORTING INFORMATION FOR
# Mott switching and structural transition in the metal phase of VO₂ nanodomain

**Materials and Methods**

Synthesis of VO$_2$

Thin films of VO$_2$ were grown on ($\bar{1}$102) (*r*-plane) Al$_2$O$_3$ substrate by a pulsed laser deposition method using a 248 nm KrF excimer laser. During the films growth, a metallic vanadium target was ablated by laser radiation (pulse energy – 300 mJ, repetition rate – 10 Hz) in a high-vacuum chamber that was filled with an oxygen gas and carefully maintained at a constant partial pressure of 30 mTorr. The substrates were kept at a temperature of 600 °C to ensure a high crystallinity of the films.

Device fabrication. Resistance and X-ray diffraction measurements

At the four corners of the film (5 mm x 5 mm), gold pads were deposited as electrical contacts for resistivity measurements. The sample temperature was controlled by heating/cooling stages from Linkam (UK). The sample was mounted to the stage with silver paste, and a K-type thermocouple was inserted to the sample stage and secured with silver paste. Simultaneous resistivity measurements were performed through gold pads at the sample corners. Further X-ray diffraction measurements were performed at HXMA beamline at the Canadian Light Source. Incident X-ray energies were selected with the Si(111) fixed-exit double-crystal monochromator. To characterize the crystalline structure (Fig. 1b), X-ray diffraction peaks were measured with broad accepting angle ranges of 3 × 6 degrees. The assignment of diffraction peaks is shown as stereographic schematics in Fig. S1. It has been reported that (011)$_M$ diffraction peaks from twin domains of (200)$_M$ and ($\bar{2}$11)$_M$ are aligned along the same substrate directions.[1] The two (011)$_M$ peaks in the Figs. 1b, 1c are originated from the two twin domains. The (11$\bar{1}$)$_M$ peak is originated from a (200)$_M$ domain with different stacking sequence from that producing the (011)$_M$. The (011)$_M$ peak changes to (101)$_R$, and (11$\bar{1}$)$_M$ peak disappears above the transition temperature.

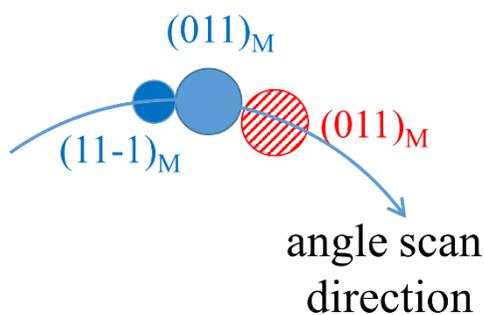

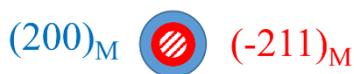

**Figure S1**. Stereographic schematics showing diffraction peak assignment in Figs. 1b, c. The (200)$_M$ and ($\bar{2}$11)$_M$ are film normal directions and the same filled pattern (or color) implied

the peaks originated from the same domain. The arrow shows angle scan direction used in the Figs. 1b, c.

The monoclinic phase lattice parameters obtained by the X-ray diffraction measurements are $a = 5.74$ Å, $b = 4.52$ Å, $c = 5.38$ Å, $\alpha = \gamma = 90°$, $\beta = 122.6°$. $d$-spacings are 3.31085 Å for $(11\bar{1})_M$, 3.20049 Å for $(011)_M$, and 2.41784 Å for $(200)_M$, 2.42282 the Å for $(\bar{2}11)_M$. The separation of $(200)_M$ and $(\bar{2}11)_M$ is difficult. Rutile phase lattice parameters are $a = b = 4.555$ Å, $c = 2.851$ Å, $\alpha = \beta = \gamma = 90°$. $d$-spacing is 2.41685 Å for $(101)_R$, which is similar to those of $(200)_M/(\bar{2}11)_M$. $d$-spacing of $(110)_R$ is 3.22059, which is similar to that of $(011)_M$.

### DANES setup

Diffraction anomalous near-edge structure (DANES) experiments using a synchrotron light source were performed by X-ray generated from a 2.1 Tesla wiggler in synchrotron at the Canadian Light Source (Fig. S2).

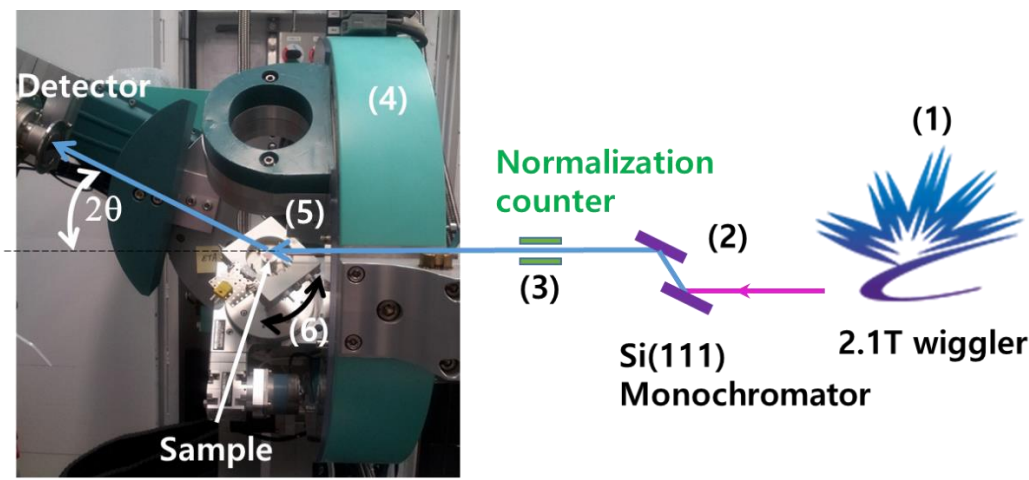

**Figure S2.** DANES experimental setup. Incident X-ray from a 2.1 Tesla wiggler (1) in synchrotron (Canadian Light Source) and a Si (111) monochromator (2) for the energy selection. Photon counts from the ionization chamber (3) in front of the sample were used to normalize the diffracted X-ray intensities. The Huber diffractometer (4) was used to find Bragg diffraction peaks and to scan sample angles. The sample temperature was controlled by the Linkam stage (5). Diffracted intensities were measured with a detector mounted on a two theta ($2\theta$) arm of the diffractometer, while the sample was moved around the sample normal direction (6).

### Θ-2θ XRD measurements

Fig. S3 shows the room temperature $\theta$-$2\theta$ XRD pattern of the epitaxial $VO_2$ film with $(200)/(\bar{2}11)$ out-of-plane orientation grown on a single-crystalline $r$-$Al_2O_3$ substrate. The measurements were performed by a high-resolution X-ray diffractometer (PANalytical) with Cu K$\alpha$1 radiation. The $VO_2$ peak at $2\theta \approx 36.5°$ ambiguously originated from $(200)$ or/and $(\bar{2}11)$ atomic planes due to their similar Bragg angles.

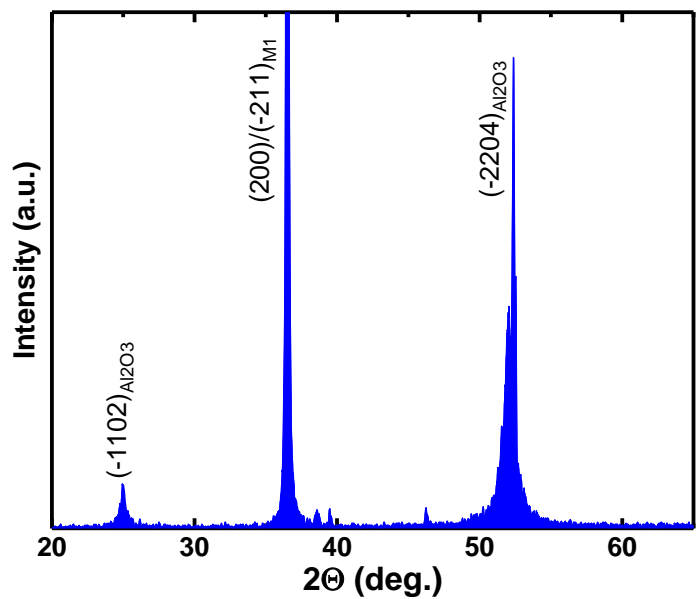

**Figure S3.** Room temperature θ-2θ XRD pattern of the VO$_2$/Al$_2$O$_3$ sample.

**VO$_2$ Domain Size Calculations**

In Fig. 3b of the main text, we demonstrate the changes in the monoclinic domain size of VO$_2$ on heating. Here, we show the corresponding calculation details. In particular, the monoclinic domain sizes were estimated from the broadening of the $(11\bar{1})_M$ VO$_2$ diffraction peak (the left peak shown in Figs. 1b,c and its width extracted in Fig.1d) by applying the Scherrer formula[2]:

$$D = \frac{K\lambda}{\beta \cos\theta} \quad (S1),$$

where $D$ is the domain size,
$K$ is the Scherrer constant ($K = 0.94$ for spherical domains),
$\lambda$ is the X-ray wavelength ($\lambda = 2.1752$ Å),
$\beta$ is the FWHM of $(11\bar{1})_M$ VO$_2$ peak,
$\theta$ is the Bragg angle of $(11\bar{1})_M$ ($\theta = 19.1775°$).

**Table S1.** The temperature dependent values of the FWHM of the $(11\bar{1})_M$ VO$_2$ diffraction peak and the corresponding domain size calculated by Eq. (S1).

| Temp. ($T$,°C) | $(11\bar{1})_M$ FWHM ($\beta$, deg.) | Domain size ($D$, Å) |
| --- | --- | --- |
| 22.562 | 0.644±0.008 | 192.505±4.493 |
| 28.043 | 0.643±0.008 | 192.888±4.436 |
| 36.234 | 0.644±0.008 | 192.600±4.474 |
| 43.459 | 0.641±0.008 | 191.977±4.480 |
| 46.744 | 0.645±0.008 | 192.321±4.488 |
| 51.016 | 0.643±0.008 | 192.900±4.605 |
| 55.617 | 0.634±0.008 | 195.689±4.774 |
| 59.598 | 0.644±0.010 | 192.756±5.718 |
| 61.357 | 0.687±0.013 | 180.459±6.775 |
| 62.62 | 0.821±0.017 | 151.071±6.007 |
| 64.325 | 1.001±0.028 | 123.915±6.772 |
| 65.312 | 1.135±0.033 | 109.269±6.125 |
| 66.679 | 1.247±0.057 | 99.451±8.755 |
| 67.912 | 1.412±0.105 | 87.870±12.187 |
| 68.796 | 1.445±0.148 | 85.822±15.973 |

**Extraction of the Band-Filling Factor $\rho$ from the Resistance Data Using the Inhomogeneous Model**

Physical meaning of $\rho$

We consider the insulator-metal-transition resistance measured in a correlated system with both the insulator and metal regions (inset of Fig. S4a), such as the temperature dependence of the resistance measured in VO$_2$ (Fig. S4a, also shown in Fig.1a). In order to find the extent of inhomogeneity, we use the model[(8)-(7)] to extract the effective mass in the inhomogeneous system.

In the model, the band-filling factor (or normalized carrier density) $\rho$ is defined by a ratio of the number of carriers in metal, $n_{carrier}$, to the number of lattices, $n_L$ or $n_{tot}$, in the measurement region; $\rho = n_{carrier}/n_L$ or $n_{carrier}/n_{tot}$.[(5)-(7)] The physical meaning of $\rho$ is the quantitative filling of the band in k-space and the extent of the metal region in real space with both the metal and insulator phases in the measurement region. 1-$\rho$ is regarded as the size of the insulator region (see the inset of Fig. S4a).

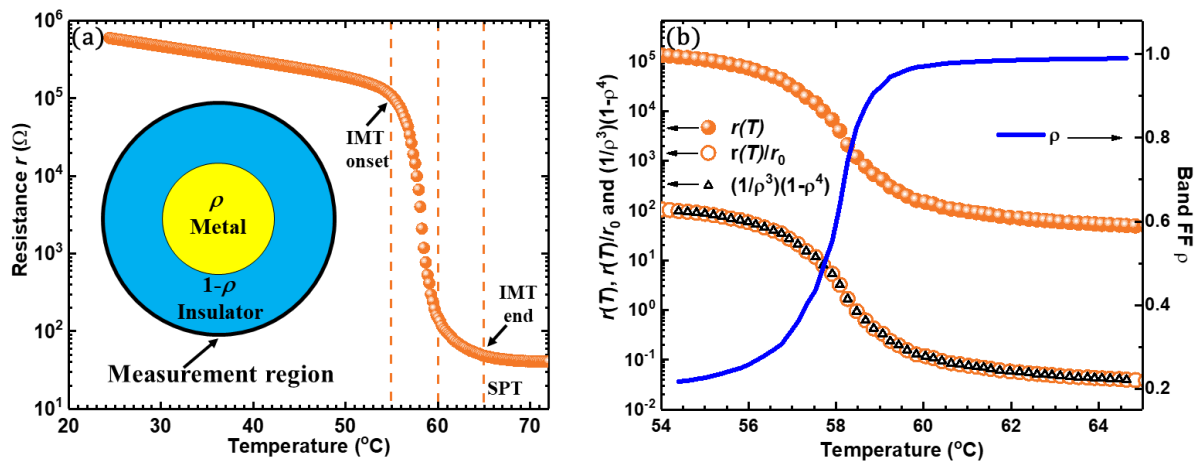

**Figure S4.** Extracting of the band-filling factor $\rho$ from the IMT resistance. (a) Resistance vs. temperature $r(T)$ of $VO_2$. The inset schematically shows a concept of a mixed-phase system with metal and insulator phases in the measurement region. $\rho$ defines the extent of the metal phase and the 1-$\rho$ insulating phase. (b) Between the insulator-metal transition (IMT) onset $T_c >54°C$ and the structural phase transition (SPT) onset $T >60°C$ in Fig S4a, the temperature dependence of resistance $r(T)$ (solid orange balls), ratio $r(T)/r_0 \equiv r(T)/1250$ (open orange balls) that is closely fitted by the derived $(1/\rho^3)(1-\rho^4)$ (open black triangles), and $\rho$ (solid blue line) are shown. The exact values of $\rho$ vs $T$ shown in Fig. 3(a) in the main text are also shown in Table S2.

**Table S2.** The temperature dependence of the resistance $r(T)$, ratio $r(T)/r_0 \equiv r(T)/1250\Omega$, the band-filling factor $\rho$ and ratio, $(1-\rho^4)/\rho^3$. $T_{SPT}$ is regarded as 64.6°C.

| $T$, °C | $r(T)$, Ohm | $r(T)/r_0$ ($r_0$=1250) | $\rho$ | $(1-\rho^4)/\rho^3$ | $T$, °C | $r(T)$, Ohm | $r(T)/r_0$ ($r_0$=1250) | $\rho$ | $(1-\rho^4)/\rho^3$ |
|---|---|---|---|---|---|---|---|---|---|
| 54.40 | 122222.00 | 97.78 | 0.22 | 97.65 | 59.62 | 194.46 | 0.16 | 0.96 | 0.16 |
| 54.61 | 118070.40 | 94.46 | 0.22 | 94.99 | 59.83 | 160.50 | 0.13 | 0.97 | 0.13 |
| 54.81 | 113052.20 | 90.44 | 0.22 | 89.95 | 60.04 | 147.84 | 0.12 | 0.97 | 0.12 |
| 55.00 | 107615.70 | 86.09 | 0.23 | 86.41 | 60.22 | 135.33 | 0.11 | 0.97 | 0.11 |
| 55.20 | 101285.90 | 81.03 | 0.23 | 80.90 | 60.42 | 117.62 | 0.09 | 0.98 | 0.10 |
| 55.40 | 94439.30 | 75.55 | 0.24 | 75.84 | 60.63 | 107.39 | 0.09 | 0.98 | 0.09 |
| 55.60 | 86359.27 | 69.09 | 0.24 | 69.45 | 60.83 | 101.64 | 0.08 | 0.98 | 0.08 |
| 55.81 | 80158.13 | 64.13 | 0.25 | 64.53 | 61.03 | 94.88 | 0.08 | 0.98 | 0.08 |
| 55.99 | 73076.91 | 58.46 | 0.26 | 58.65 | 61.21 | 90.61 | 0.07 | 0.98 | 0.07 |
| 56.17 | 65367.31 | 52.29 | 0.27 | 52.27 | 61.37 | 86.37 | 0.07 | 0.98 | 0.07 |
| 56.36 | 57012.00 | 45.61 | 0.28 | 45.77 | 61.53 | 82.10 | 0.07 | 0.98 | 0.07 |
| 56.57 | 49363.25 | 39.49 | 0.29 | 39.46 | 61.74 | 76.98 | 0.06 | 0.99 | 0.06 |
| 56.75 | 43032.40 | 34.43 | 0.31 | 34.59 | 61.97 | 72.83 | 0.06 | 0.99 | 0.06 |
| 56.93 | 33855.65 | 27.08 | 0.33 | 27.24 | 62.19 | 70.18 | 0.06 | 0.99 | 0.06 |
| 57.13 | 26414.07 | 21.13 | 0.36 | 21.07 | 62.40 | 67.14 | 0.05 | 0.99 | 0.06 |

| 57.32 | 19100.73 | 15.28 | 0.40 | 15.23 | 62.59 | 64.98 | 0.05 | 0.99 | 0.05 |
| 57.53 | 14525.78 | 11.62 | 0.44 | 11.63 | 62.77 | 63.12 | 0.05 | 0.99 | 0.05 |
| 57.71 | 9959.88 | 7.97 | 0.49 | 7.96 | 62.93 | 61.42 | 0.05 | 0.99 | 0.05 |
| 57.91 | 6676.09 | 5.34 | 0.55 | 5.34 | 63.12 | 59.36 | 0.05 | 0.99 | 0.05 |
| 58.08 | 4008.59 | 3.21 | 0.64 | 3.21 | 63.33 | 57.63 | 0.05 | 0.99 | 0.05 |
| 58.26 | 2094.40 | 1.68 | 0.75 | 1.67 | 63.53 | 56.01 | 0.04 | 0.99 | 0.04 |
| 58.47 | 1185.91 | 0.95 | 0.83 | 0.95 | 63.74 | 55.20 | 0.04 | 0.99 | 0.04 |
| 58.67 | 772.40 | 0.62 | 0.88 | 0.62 | 63.90 | 53.97 | 0.04 | 0.99 | 0.04 |
| 58.86 | 532.78 | 0.43 | 0.91 | 0.43 | 64.07 | 52.94 | 0.04 | 0.99 | 0.04 |
| 59.06 | 416.76 | 0.33 | 0.93 | 0.33 | 64.26 | 51.92 | 0.04 | 0.99 | 0.04 |
| 59.24 | 299.41 | 0.24 | 0.95 | 0.24 | 64.47 | 51.12 | 0.04 | 0.99 | 0.04 |
| 59.43 | 244.45 | 0.20 | 0.95 | 0.20 | 64.63 | 49.94 | 0.04 | 0.99 | 0.04 |

**Derivation of the metal resistance from Drude model**

The resistivity of metal in the Drude picture is defined at one electron per atom as:

$$r = \frac{m}{ne^2\tau}, \quad (S2)$$

where $m$ is the band mass of a carrier, $n$ is the carriers' density ($n \equiv n_{carrier}$), $e$ is the elementary charge, and $\tau$ is the scattering time between carriers (assumed to be constant).
We introduce an inhomogeneous (mixed phase) correlated system. Then, $m$ is replaced by the true correlated effective mass of a carrier, $m^*$. In the inhomogeneous model,[(5)-(7)] the element charge is defined as the effective charge as follows,

$e^* = \rho e$                (S3),

the effective mass is given as

$m^* = m/(1-\rho^4)$,       (S4),

$n$ is also denoted as

$n \equiv n_{carrier} = \rho n_{total}$       (S5).

In the case of a strongly correlated metal, because the resistivity and the effective mass should be maximum, $m$ in Eq. (S2) is the true band mass, which is replaced by $m \equiv m^*(1-\rho^4)$ where $m^*$ is in the BR picture.[(29)] This decreases with increasing $\rho$.
Accordingly, when Eq. (S3), Eq. (S4) and Eq. (S5) are applied to Eq. (S2),
Eq. (S2) is newly given as follows:

$$r = \frac{m}{n_{carrier}e^{*2}\tau} = \frac{m^*(1-\rho^4)}{n_{tot}e^2\tau\rho^3},$$

$$= r_0 \left[\frac{1-\rho^4}{\rho^3}\right], \quad (S6)$$

Where $r_0 = (m^*/n_{tot}e^2\tau)$ is constant, which is defined as the Drude resistivity.
Thus, the ratio of resistivity $Ratio \equiv r(T)/r_0$ is given as follows (Fig. S5):

$$\frac{r(T)}{r_0} = \left[\frac{1-\rho^4}{\rho^3}\right], \quad (S7)$$

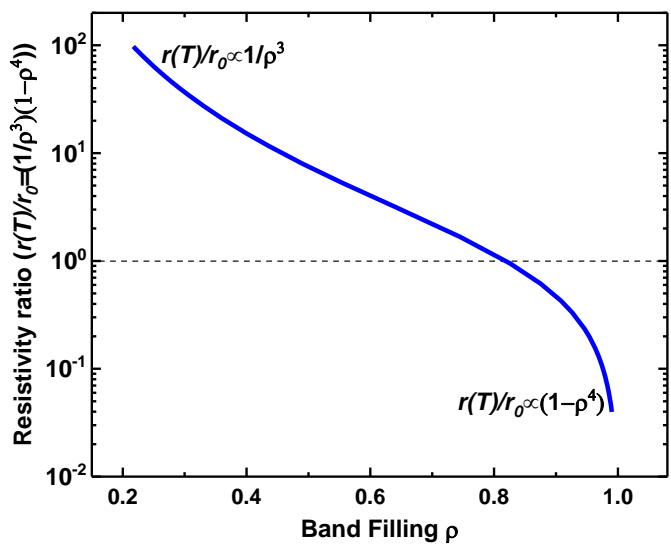

**Figure S5.** The dependency of the resistivity ratio $r(T)/r_0=(1-\rho^4)/\rho^3$ upon the band-filling factor $\rho$ and $r_0$=1,250. Note that $r(T)/r_0 \to 0$ at $\rho \to 1$ represents the transition to the metallic state. At a low $\rho$, $r(T)/r_0$ is proportional to the $1/\rho^3$ coming from the Drude picture and, at a high $\rho$, $r(T)/r_0$ is proportional to $(1-\rho^4)$ attributed to the effective mass.

In order to find the Drude resistance, $r_0$, we give a boundary condition of the metal range between $T_{onset}$ and $T_{SPT}$. At $T_{onset}$=54.6°C, the resistance is 118,070Ω and at $T_{SPT}$=64.6°C the resistance is 50Ω (Table S2). Near $T_{SPT}$, $\rho$ can be assumed as $\rho \approx 0.99$, 0.999, or 0.9999 of approximately 1. When an appropriate $\rho$=0.99 is selected, $r(T)/r_0=(1-\rho^4)/\rho^3 \approx 0.0399 \approx 0.04$ is obtained, and, by applying $r(T) \approx 50Ω$ to $r(T)/r_0 \approx 0.04$, $r_0$ is determined as 1250Ω. Since solving Eq. S6 is difficult, numerical analysis is recommended.

Thus, Eq. (S6) is re-expressed as follows:

$$\text{Ratio} \equiv \frac{r(T)}{1250} = \frac{1-\rho^4}{\rho^3}, \quad (S8)$$

which is given in Fig. S4b.

Then, when $r(T)$ is between $T_{onset}$ and $T_{SPT}$ at Fig. S4b and is applied to Eq. (S8), $\rho$ is extracted from the $r(T)/r_0$. The $\rho$ extracted by applying Fig. S5 from the metal range of the resistance data is shown in Fig. S4(b).

**Analysis of the "Mottian" Characteristics of $VO_2$**

The observation of the monoclinic metal phase (MMP) in the manuscript (Figs. 1d, 2 and 3a,b) is strong evidence of $VO_2$ being the Mott insulator. One of the features of the Mott-type IMT is the divergent effective mass defined in the extended Brinkman-Rice picture[5]-[7] as follows:

$$\frac{m^*}{m} = \frac{1}{1-\left[\frac{U}{U_c}\right]^2} = \frac{1}{1-\kappa_{BR}^2 \rho^4}, \qquad (S9)$$

where $U$ is the on-site Coulomb energy, $U_c$ is the critical on-site Coulomb energy, $\kappa_{BR}$ is the correlation strength,[5] and $\rho$ is the band-filling factor.

We find the correlation-strength value of $\kappa_{BR}$ at the MMP near the structural phase transition edge of 65°C. $m^*/m = 25.4$ at $\rho \approx 0.99$ (Table S2) and $\kappa_{BR} = 1$ in Eq. (S9) is determined. When $\rho = 1$ at Eq. (S9) is approximately assumed, thus, we obtain from $m^*/m = 25.4$:

$$\kappa_{BR} = \sqrt{1-\frac{m}{m^*}} \approx 0.98.$$

## References

(1) Yang, T.-H.; Aggarwal, R.; Gupta, A.; Zhou, H.; Narayan, R.J.; Narayan, J. Semiconductor-metal transition characteristics of VO2 thin films grown on c- and r-sapphire substrates. *Journal of Applied Physics* **2010**, 107, 053514.

(2) Patterson, A. L. The Scherrer Formula for X-Ray Particle Size Determination. *Physical Review* **1939**, 56, 978.

(3) Kim, H.-T.; Chae, B.-G.; Youn, D.-H.; Maeng, S.-L.; Kim, G.; Kang, K.-Y.; Lim, Y.-S. Mechanism and observation of Mott transition in VO$_2$-based two- and three-terminal devices. *New Journal of Physics* **2004**, 6, 52.

(4) Kim, H. T. Analysis of the diverging effective mass in YaBa$_2$Cu$_3$O$_{6+x}$ for high-$T_c$ mechanism and pairing symmetry, *International Journal of Modern Physics* B **2018**, 32, 1840031.

(5) Kim, H. T. Extension of the Brinkman-Rice picture and the Mott transition, *Physica C: Superconductivity* **2000**, 341-348, 259-260. (https://arxiv.org/pdf/cond-mat/0001008v6.pdf).

(6) Kim, H. T. Extended Brinkman-Rice Picture and Its Application to High-Tc Superconductors. http://arXiv.org/abs/cond-mat/0110112.

(7) Kim, H. T. Fitting of *m*/*m* with Divergence Curve for He$_3$ Fluid Monolayer using Hole-driven Mott Transition. APS March Meeting **2012**, Abstract No: A16.00015.

(8) Brinkman, W. F.; Rice, T. M. Application of Gutzwiller's variational Method to the Metal-Insulator Transition, *Physical Review* B **1970**, 2, 4302.